\documentclass[%
aip,
sd,%
amsmath,amssymb,
reprint,%
]{revtex4-1}
\usepackage[colorlinks=true,
linkcolor=red,
urlcolor=black,
citecolor=blue]{hyperref}
\usepackage{graphicx}
\usepackage{epstopdf}
\usepackage{dcolumn}
\usepackage{bm}
\begin{document}
\preprint{AIP/123-QED}
\title{Alternating chimeras in networks of ephaptically coupled bursting neurons}
	\author{Soumen Majhi}
	\affiliation{$^1$Physics and Applied Mathematics Unit, Indian Statistical Institute, 203 B. T. Road, Kolkata-700108, India}

	\author{Dibakar Ghosh}\email{diba.ghosh@gmail.com}
	\affiliation{$^1$Physics and Applied Mathematics Unit, Indian Statistical Institute, 203 B. T. Road, Kolkata-700108, India}
	\date{\today}

	\begin{abstract}
		The distinctive phenomenon of chimera state has been explored in neuronal systems under a variety of different network topologies during the last decade. Nevertheless, in all the works, the neurons are presumed to interact with each other directly with the help of synapses only. But the influence	of ephaptic coupling, particularly magnetic flux across the membrane is mostly unexplored and should essentially be dealt with during the emergence of collective electrical	activities and propagation of signals among the neurons in a network. Through this article, we report the development of an emerging dynamical state, namely, the {\it alternating chimera}, in a network of identical neuronal systems induced by an external electromagnetic field. Owing to this interaction scenario, the nonlinear neuronal oscillators are coupled indirectly via electromagnetic induction with magnetic flux, through which neurons communicate in spite of the absent physical connections among them. The evolution of each neuron, here, is described by the three-dimensional Hindmarsh-Rose dynamics. We demonstrate that the presence of such non-locally and globally interacting external environments induce a stationary alternating chimera pattern in the ensemble of neurons, whereas in the local coupling limit the network exhibits transient chimera state whenever the local dynamics of the neurons is of chaotic square-wave bursting type. For periodic square-wave bursting of the neurons, similar qualitative phenomenon has been witnessed with the exception of the disappearance of cluster states for non-local and global interactions. Besides these observations, we advance our work while providing confirmation of the findings for neuronal ensembles exhibiting plateau bursting dynamics and also put forward the fact that plateau pattern actually favors the alternating chimera more than others. These results may deliver better interpretations for different aspects of synchronization appearing in network of neurons through field coupling that also relaxes the prerequisite of synaptic connectivity for realizing chimera state in neuronal networks.

	\end{abstract}
	
	\maketitle
	\begin{quotation}
		{\bf Neural networks, among other complex systems, self-organize in such ways that synchronous spatiotemporal patterns	may appear. Different aspects of synchronization are extremely fundamental neural mechanisms. They assist in neural communication, neural plasticity and are  important for many cognitive processes. Chimera-like patterns that deal with co-existence of synchronized and de-synchronized domains in the same system, bear a strong resemblance to several neuronal developments. The observation of such chimera state in neuronal systems includes several notable works, all of which contemplate with electrical or chemical (or both) synapses as the communicating medium among the neurons. But, the effects of electromagnetic induction must not be neglected at the time of fluctuation in inter- and extra-cellular ion concentrations. In this work, we propose a network model of neurons exposed to external electromagnetic field that in turn is responsible for communication among the neurons, in absence of any kind of synaptic interactions among them. Noticeably, we encounter the alternating chimera together with the transient chimera pattern in the network depending on the coupling radius. The lifetime of transient chimera patterns may differ for the cases of chaotic and periodic bursting of the local dynamical systems. In addition, cluster states (for chaotic bursting of the neurons) along with multicluster oscillation death have been realized in the network depending upon the coupling radius and interaction strength. Moreover, the occurrence of all the dynamical phenomena are confirmed for neuronal network based upon plateau bursting dynamics that has been witnessed to broaden the alternating chimera region in the parameter space.

		  }
		
	\end{quotation}

\section{Introduction}
The brain and nervous system are amazing complex structures whose activities are modeled based on the interactions among the neurons. The presence of billions of neurons and the diverse interaction patterns among them make the cortical network possibly the most challenging complex system.
Most of the cognitive functions in brain are based on the
coordinated activities in large numbers of neurons
distributed over specialized brain areas. Various forms of synchronization in neural oscillations (in low and high frequencies) are quite essential components for facilitating coordinated activity in the normally functioning brain. On the other hand, in nonlinear dynamics literature, exceptional spatial concurrence of coherent and incoherent dynamical behaviors arising in network of coupled oscillatory systems is popularly known as the {\it chimera state} \cite{chim_rev1, chim_rev2}. The recognition of such a captivating collective phenomenon was initiated with Kuramoto's observation in a nonlocally coupled system of identical phase oscillators \cite{chp1}, since then it has brought a broad research field in the literature of nonlinear dynamics. It has
been well-established that these are not limited to network of phase oscillators \cite{chp1,chp2,chp3,chp5,chp6}, rather this unique collective state can also appear in a large variety of other systems \cite{chco1,chco2,chhn1,chhn3,chhn4} including neural networks \cite{chne1,chne2,chne4,chne5,chne6,chne7,tanmoy,chne8,chnen,ch2d}.
Both regular symmetric topology (local, non-local and global) \cite{global,chne7,chlo1,tanmoy} together with anomalistic interactions \cite{chne4,chne6,chcom,chtv,chyao,chne8,chnen} on top of networks have been confronted so far in order to realize chimera-like patterns. Appearance of several variants of the chimera patterns, such as globally clustered chimera \cite{laksh2}, multi-chimera \cite{chne1}, traveling chimera \cite{tanmoy}, breathing chimera \cite{chp3}, amplitude mediated chimera \cite{amc}, imperfect chimera \cite{imc}, virtual chimera \cite{chexp6} etc. are reported. Besides these, the emergence of chimeras have been identified in multilayer networks \cite{chne8,chnen,mul1,chmult2,chmult3,mul2} and also been revealed experimentally \cite{chexp6,chexp2,chexp3}.
\par Since the time of its detection, the chimera-like patterns have been strongly connected to various neuronal activities, such as the bump states in neural systems \cite{bu1,bu2}, the real world phenomena of unihemispheric slow-wave sleep \cite{sleep1,sleep2} of some aquatic animals (e.g. dolphins, eared seals) and of some migrated birds. This also includes
various types of pathological brain states \cite{braindis1, braindis2} such as the Alzheimer's disease, epilepsy, autism, schizophrenia and brain tumors. Due to the existence of such definite correspondences between the chimera-like patterns and several neuronal evolutions, in recent times there have been efforts \cite{chne1,chne2,chne4,chne6,chne7,tanmoy,chne8,chnen,ch2d} to study the emergence of such unexpected patterns in neuronal networks with different views of interactional form.
\par Nevertheless, the articles noted above deliberately thought of the synapses (electrical or chemical or both) as the  only communicating (information transferring) medium in the respective considered neuronal networks. Of course, synapses are essential for neuronal functions, that allow a neuron to pass an electrical or chemical signal to another neuron. But, {\it what if the neurons in the network are not coupled through synapses?} This is a question of severe importance and has been a topic of discussion for decades. Researches concerning this issue attest to the fact that answer to this can be given through the mechanism of the {\it ephaptic communication}.
It may correspond to the interaction of nerve fibers in contact due to the exchange of ions between the cells. On the other hand, it refers to the coupling of nerve fibers because of the extracellular local electric fields. Illustrative discussions \cite{eph1,eph2,eph3,eph4,eph5,eph6,eph7,eph8,eph9,danko,nlin1,nlin2,nlin3} exist on the issue of non-synaptic ephaptic communication among neurons within the nervous system based on the latter one, that essentially differs from the direct communication processes through electrical and chemical synapses.  These articles indicate that the external fields carried by the extracellular medium do plenty of works and that they may, actually, represent an additional form of neuronal communication.

\par  Moreover, recent researches \cite{eph3,eph7,eph9,nlin2,neurocom} suggest that the modeling of neuronal networks as mere an ensemble of physical connections (biochemical or electrical) among neurons is incomplete unless it also includes effusive extra-neuronal phenomena such as electric and magnetic fields. Indeed, the dynamics of the neuronal activity may get affected because of the fluctuation in the inter- and extra- cellular ion concentrations. Consequently, internal variation in the electromagnetic field may develop and hence the influence of magnetic flux across the membrane must be taken into account in order to analyze information transferring and collective electrical activities in neuronal ensembles. Interestingly, the electromagnetic fields thus created through neuronal activities can send signals and information to neighboring neurons without following any synaptic information exchange procedures \cite{danko}. So, it will be of great worth studying emergence of diverse collective behaviors arising in network of neurons communicating through the external electromagnetic fields. Previously, this raised issue has been dealt with in a few recent works on the basis of origination of possible collective synchrony amongst a few neurons \cite{nlin1,nlin2,nlin3}.
As far as the dynamical consequences in a network of indirectly coupled neurons is concerned, one of the previous works suggest that the external common noisy field \cite{noi} can have a positive influence in the context of inducing or enhancing synchronization among the neurons. Neuron's membrane potential stimulation \cite{sti} can also be a good option regarding this issue of bringing synchrony. The works \cite{chne8,chnen} unlocked the appearance of chimera states in network of uncoupled neurons induced by a multilayer formalism, of course, dealing with synapses (electrical and chemical) after all.
\par At the same time we would like to note that previously, alternating chimeras in which coherent and incoherent domains alternate their spatial positions and hence in the level of synchrony over time, have been realized in only a few works. For instance, in time-static networks of two phase oscillator populations \cite{chaltepl,chaltpre} with time delays and another one in time-varying network \cite{chtv}, alternating chimera has been observed. This pattern is recognized in an oscillatory medium with nonlinear uniform global coupling \cite{chaltsrep}. But this peculiar temporal alternating nature of chimera state that explains the phenomenon of dynamic unihemispheric alternating sleep the best, is still mostly unexplored and yet to be given its due attention. In the present work, we unravel this unique dynamical phenomenon of the alternating chimera patterns in a neuronal network with the neurons interacting via external electromagnetic field. Communication among the neurons through this indirect ephaptic formalism can be modeled locally, non-locally or globally. We have gone through all these topologies while considering three-dimensional Hindmarsh-Rose models as the local dynamical systems. Chaotic square-wave bursting dynamics of the systems under the above explained configuration with non-local or global limit has been witnessed to produce stationary alternating chimera patterns as a link between incoherence and coherent (or cluster) states followed by the multi-cluster oscillation death states. In the limit of local coupling, transient chimera (leading to fully disordered state over time) is also realized. Whenever the dynamical units of the network follow periodic bursting, the transient chimera patterns may be observed to have higher lifetimes \cite{lft} but the network does not go through any cluster-like states. Furthermore, prominence in our results is substantiated by providing evidence in case of plateau bursting of the neurons, which makes the proposed mechanism of ephaptic communication for realizing the alternating chimera pattern, quite general. Significantly enough, the proposed mechanism thus softens the fundamental synaptic connectivity requirement for realizing chimera state in neuronal networks.

\par The remaining part of this paper is organized as follows. In Sec. II, we discuss the mathematical model of the network considered here. Sec. IIIA  illustrates the case of emergence of the stationary alternating chimera in interacting chaotic bursting Hindmarsh-Rose models. In Sec. IIIB, we report how alternating (and transient) chimera may appear for periodic bursting of the neurons in the network followed by a discussion on confirmation of the obtained results for neurons possessing plateau bursting dynamics in Sec. IIIC. Finally, Sec. IV provides the concluding remarks.

\section{ Mathematical model}
\par  This section is devoted to the mathematical description of the considered neuronal network model. The effect of external electromagnetic field through the introduction of magnetic flux is presented, where Hindmarsh-Rose dynamics is used to cast each node, as follows:\\

\begin{equation}
\begin{array}{lcl}\label{eq1}
\dot x_i=y_i+b{x_i}^2-a{x_i}^3-z_i+I-\epsilon\rho(\phi_i)x_i,\\
\dot y_i=\alpha-d{x_i}^2-y_i,\\
\dot z_i=c[s(x_i-e)-z_i],\\
\dot \phi_i=-k_1\phi_i+k_2 x_i+\sum\limits_{j=i-P}^{j=i+P}(\phi_j-\phi_i),
\end{array}
\end{equation}
where $(x_i,y_i,z_i)$ ($i=1, 2, \cdot \cdot \cdot ,N$) represent the state vectors for the nodes, particularly variables $x_i$ represent the membrane potentials of the neurons and the variables $y_i$ and $z_i$ correspond to the transport of ions across the membrane via the fast (associated to $\mbox{Na}^+$ or $\mbox{K}^+$) and slow (associated to $\mbox{Ca}^{2+}$) channels. Here $N$ is the number of neurons in the network and the parameters $b=3.0$, $a=1.0$, $\alpha=1.0$, $d=5.0$, $s=4.0$, $e=-1.6$ and the slow dynamics modulator $c=0.005$ are chosen so that with the external forcing current $I=3.25$ and $I=1.90$, the individual neurons display multi-scale chaotic bursting and periodic bursting respectively. Neuronal bursting is extremely important for neuronal communication, particularly for motor pattern generation, synchronization etc. that facilitates neuro-transmitter release, also helps in overcoming synaptic transmission failure.
\par As pointed out above, due to the fluctuation in ion concentrations, electromagnetic induction may affect electrical activities of the neurons. In fact, density of magnetic flux across membrane gets changed when the
neurons are exposed to electromagnetic field that assists in communication and is also capable of defining the memory effect in neurons. In this connection, we should mention that the concept of time delay can be used to explain effect of memory (while making the system infinite dimensional) in neurons as well. But magnetic field can also be very efficient to illustrate the impact of memory in neurons while generating a proper spatial distribution and consequently the fluctuation of electromagnetic field makes information exchange possible. Furthermore, neurons are also considered as intelligent circuits dealing with complex signals in the nervous system. So the functionality of a proper memristive system can resemble the synaptic interactions in neuronal networks and memristors \cite{memr1} are nonlinear electrical components regulating the current flow in a circuit that links electric charge and magnetic flux  and remembers the amount of charge that has previously flowed through it. Thus as far as the communication
among the neurons is concerned, here (cf. Eq. (\ref{eq1})) $\phi_i$ defines the magnetic flux across the membrane and $\rho(\phi_i)$ describes the memory conductance (memductance) of a magnetic flux-controlled memristor that governs the coupling between the flux and membrane potential of neurons with $\epsilon$, $k_2$ being the strengths controlling the interaction. This memductance is often represented \cite{nlin1,nlin2,nlin3,neurocom,memr2,memr3} as\\
\begin{equation}
\rho(\phi_i)=\beta_1+3\beta_2{\phi_i}^2,
\end{equation}
$\beta_1$, $\beta_2$ being fixed parameters. In fact, through the final term of the fourth equation (cf. Eq. (\ref{eq1})) the magnetic contribution of other $j=i-P,~i-P+1,\cdot \cdot \cdot,~~i+P$ neurons to the $i$-th neuron is described and hence via the magnetic flux $\phi_i$, the $i$-th neuron of the ensemble in turn interacts with $P$ neurons on both sides of an one dimensional ring that signifies an indirect non-local formalism with periodic boundary conditions. The parameter $r=P/N$ is usually termed as the coupling radius in the literature. Further, we have taken $k_1=0.5$, $k_2=0.9$ \cite{gene} and $\beta_1=0.40$, $\beta_2=0.02$, throughout this work.

\section{Results}
\begin{figure}
	\centerline{
		\includegraphics[scale=0.55]{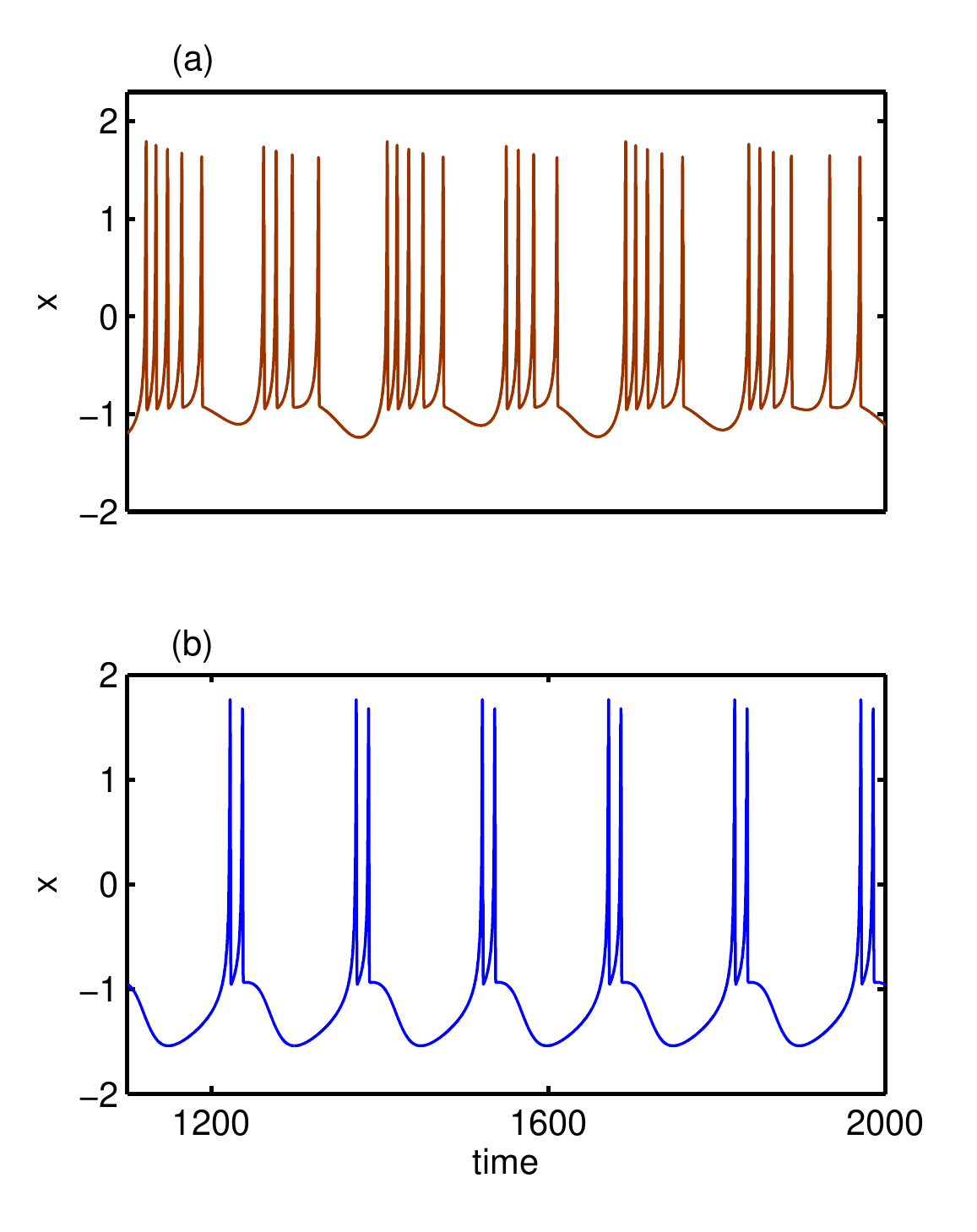}}
	\caption{(a) Chaotic square-wave bursting for $I=3.25$, (b) periodic square-wave bursting (doublet) dynamics for $I=1.90$, in one of the isolated neurons. }
	\label{tsr}
\end{figure}

Whenever there is no interaction among the neurons (i.e., for $\epsilon=0$), with $I=3.25$ (and the other parameters as stated before) neurons exhibit a typical chaotic square-wave bursting dynamics, as depicted in Fig.\ref{tsr}(a). But for $I=1.90,$ all the neurons display periodic bursting, as in Fig.\ref{tsr}(b). Here we note that in systems having directly interacting dynamical units with $P$ nearest neighbors on each side of a ring, in order to isolate the units one can set either the coupling strength $\epsilon=0$ or coupling range $ P=0$. But here, in our model (Eq. \ref{eq1}), the units (the neurons) do not really interact directly through the membrane potentials $x_i$.  Rather,  they communicate indirectly via different variables representing magnetic flux $\phi_i$ across the membrane.  Here $\epsilon$ and $k_2$ are the two parameters that respectively governs the mutual effects between membrane potential and magnetic flux and so even if one sets $P=0$ (or equivalently $r=0$), there would be terms remaining in the system that may affect the original uncoupled dynamics of the neurons because of their correlations to the magnetic flux. But still the neurons remain completely uncoupled with each other as the associated fields are not interacting then.
\par Let us now start by looking at the dynamical behavior of the neurons by changing the interaction strength $\epsilon$. For this, we will primarily concentrate on chaotic square-wave bursting dynamics of the neurons in subsection IIIA followed by its periodic counterpart in IIIB and finally plateau bursting in IIIC.

\begin{figure*}
	\centering\includegraphics[width=1.0\linewidth]{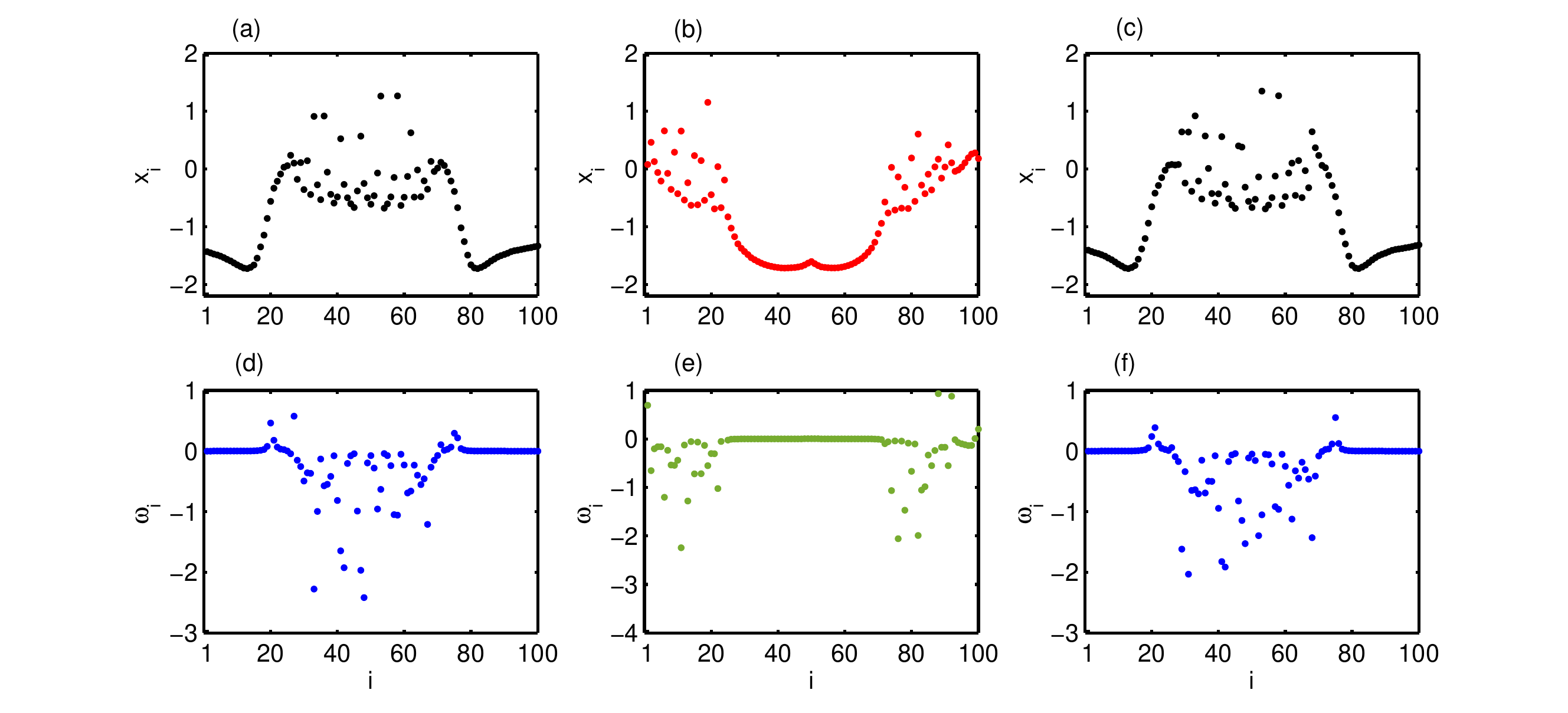}
	\caption{ Snapshot of the membrane potentials $x_i$ depicting chimera state at (a) $t=1600$, (b) $t=1700$ and (c) $t=1800$. The corresponding instantaneous angular frequencies $\omega_i$ are respectively shown in (d), (e) and (f). Here $r=0.3$.}
	\label{snapp}
\end{figure*}

\subsection{Chaotic square-wave bursting of the neurons}

\par Initially for a fixed coupling radius $r=0.3$ with communication among the neurons through $\epsilon$ turned on, the networked system (Eq. (\ref{eq1})) remains in incoherent (desynchronized) state until $\epsilon$ reaches $\epsilon=0.33$ \cite{mthdinit}. Beyond this value of $\epsilon$, the network starts realizing chimera patterns, as shown by the snapshots of the membrane potentials in Fig. \ref{snapp} where the interaction strength is $\epsilon=0.5$. Snapshot of the membrane potentials $x_i$ are shown in Figs. \ref{snapp}(a), (b) and (c) respectively at three different times $t=1600$, $t=1700$ and $t=1800$. The presence of two spatially coherent domains mediated by an incoherent domain is conspicuous from Fig. \ref{snapp}(a). But these locations of the coherent and incoherent domains get interchanged in the next snapshot (cf. Fig. \ref{snapp}(b)). However, a similar spatial arrangement in the snapshot as in Fig. \ref{snapp}(a) is again observed in Fig. \ref{snapp}(c). This indicates that there may be some sort of periodic alteration in the coherent (incoherent) domain formation of the system over time, readily signifying an alternating nature of the chimera profile. Before going into the detailed understanding of this behavior, we firstly validate the appearance of this chimera pattern by computing instantaneous angular frequencies $\omega_i$ of all the neurons as
\begin{equation}
\begin{array}{lcl}\label{eq6}
\omega_{i}= \dot{\psi}_{i} =\frac{x_{i} \dot{y}_{i}-\dot{x}_{i} y_{i}}{x_{i}^2+y_{i}^2},
\end{array}
\end{equation}
where $\psi_i(t)=\mbox{tan}^{-1}[\frac{y_i(t)}{x_i(t)}]$ is the geometric phase associated to the fast variables $x_i$ and $y_i$ of the $i$-th neuron, which is a good approximation as long as $c$ is small ($<<1$). The instantaneous angular frequencies $\omega_i$ corresponding to the snapshots of Figs. \ref{snapp}(a), \ref{snapp}(b) and \ref{snapp}(c) are respectively plotted in Figs. \ref{snapp}(d), \ref{snapp}(e) and \ref{snapp}(f). This makes the coexistence of coherence and incoherence quite clear in which coherent cluster has the same $\omega_i$ whereas in incoherent domain neurons possess different frequencies.
\par But these snapshots portray the chimera profiles only at fixed time instants. So, further in pursuance of characterizing the chimera patterns obtained through the numerical experiments in more detail, we perform the analysis of temporal evolutions of the spatial domains possessed by coherent oscillators in terms of local curvature \cite{classchim}. Local curvature at each point in space is computed while operating discrete Laplacian $l_i$ on each snapshot of the membrane potentials $x_i~(i=1, 2, . . ., N)$.
For instance, $l_i$ (and hence $L_i$) applied on a snapshot of $x_i$ at time $t$ is defined as
\begin{equation}
\begin{array}{lcl}\label{eq2}
L_i(t)=|l_i(t)|=|x_{i+1}(t)+x_{i-1}(t)-2x_i(t)|,
\end{array}
\end{equation}
smooth profile of which represents spatial coherence and
significant non-zero curvature portraying incoherence.
Now due to the dependence of amount of coherence on the individual dynamical systems and since this coherence may not be absolute complete synchrony, so one needs to go through the further estimation procedure by defining some threshold (say, $\delta_1$) based upon the maximal curvature possessed by the system. Wherefore, we define the function $f: L\rightarrow\mathbb\{0,1\}$ where $L\equiv (L_1, L_2, . . ., L_N)$, such that
\begin{equation}
f(L_i)=
\left\{
\begin{array}{c}
1,~\mbox{if}~L_i\le \delta_1\\
0,~\mbox{otherwise}\\
\end{array}. 
\right. \\
\end{equation}
 The spatial correlation measure $C_{sp}(t)$ is thus described as
\begin{equation}
\begin{array}{lcl}\label{eq3}
C_{sp}(t)=\frac{1}{N}\sum\limits_{i=1}^{N}f(L_i).
\end{array}
\end{equation}
$C_{sp}(t)$ basically reflects the spatial extent of coherence exhibited by the networked system at the time instant $t$ (with the threshold $\delta_1$ being around $1$\% of the maximal value of $L_i$). Particularly, $C_{sp}(t)=0$ refers to the situation in which none of the $L_i$'s are zero depicting desynchronized dynamical behavior. Unit value of $C_{sp}(t)$ resembles the coherent state whereas the range $0<C_{sp}(t)<1$ symbolizes the manifestation of chimera states. By definition, $C_{sp}(t)$ is, in general, time-dependent. Static coherence in the chimeric pattern will then be identified by a non-zero constant $C_{sp}(t)$ while its (non-zero, non-unit) time-varying character signifies the existence of non-static chimera.
\par Besides this spatial correlation measure $C_{sp}(t)$, in order to quantify the extent of time-correlated nodes in the network, we will be calculating the time-correlation measure $C_{tm}$ in the following way
\begin{equation}
\begin{array}{lcl}\label{eq4}
C_{tm}=\sqrt{\frac{1}{N(N-1)}\sum\limits_{i,j=1(i\neq j)}^{N}g(|\sigma_{ij}|)}.
\end{array}
\end{equation}
Here $\sigma$ corresponds to the time-correlation coefficient and the function $g$ of $|\sigma|$ is defined as\\
\begin{equation}
g(|\sigma_{ij}|)=
\left\{
\begin{array}{c}
~~1,~\mbox{if}~|\sigma_{ij}|>\delta_2\\
0,~\mbox{otherwise}\\
\end{array}. 
\right.\\
\end{equation}
 In fact, for two membrane potentials (time-series) $x_i$ and $x_j$ with $m_i,~m_j$ and $s_i,~s_j$ respectively being their temporal means and standard deviations, the pairwise correlation coefficients $\sigma_{ij}$ are defined as
\begin{equation}
\begin{array}{lcl}\label{eq5}
\sigma_{ij}=\frac{\langle(x_i-m_i)(x_j-m_j)\rangle}{s_is_j}.
\end{array}
\end{equation}
Then, $x_i$ and $x_j$ are linearly time-correlated whenever $\sigma_{ij}\simeq1$ and are anti-correlated for $\sigma_{ij}\simeq-1$ so that $C_{tm}$ serves as a time-correlation measure. We have chosen the values $\delta_1=0.04$ and $~\delta_2=0.90$ which are sufficient here in order to discriminate respectively the spatially correlated and time-correlated units.

\par As far as the evolution of this correlation measure recognizing different dynamical behaviors is concerned, the value of $C_{tm}$ must be non-zero in the regime of any non-transient chimera no matter which time-span is taken into account for the calculation of temporal means and standard deviations. Of course, one should define $\delta_2$ depending on the level of coherence observed in the system.
It does not necessarily reflect the size of coherent domain, rather for non-static (alternating) coherent clusters its values are smaller than $C_{sp}(t)$. However, for the static chimeras where no coherence is present in the incoherent domain, $C_{tm}$ yields same values as $C_{sp}(t)$. In brief, for the system in the stationary (non-transient) alternating chimera state, one has $0<C_{sp}(t)<1,~\forall t$ with an oscillatory behavior in $C_{sp}(t)$ having $C_{tm}>0$. On the other hand, the scenario: $\exists~T:~0<C_{sp}(t)<1~,\forall t<T$ and $C_{sp}(t)=0$ or $C_{sp}(t)=1,~\forall t\ge T$  is characterized as the transient chimera. Regular repetitive variations of $C_{sp}(t)$ may also signify the breathing or traveling chimera like states. But in our case, the clear occurrence of periodically alternating coherent and incoherent domains (cf. Fig. 2 and Fig. 3 ) in the chimera pattern readily implies the emergence of alternating chimera states \cite{chtv,chaltepl,chaltpre,chaltsrep}.
\par
\begin{figure}
	\centerline{
		\includegraphics[scale=0.50]{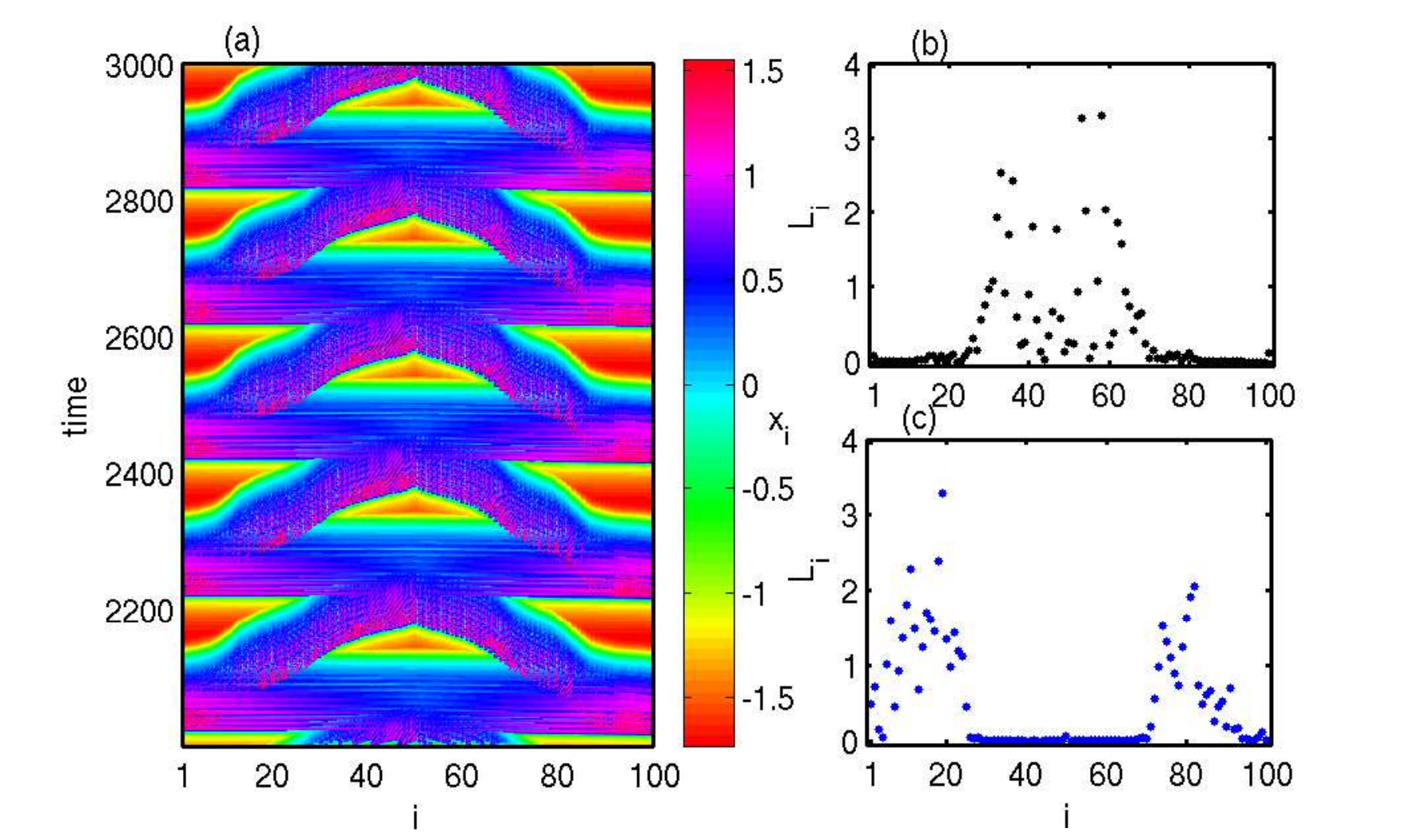}}
	\caption{ (a) Spatio-temporal plot associated to the chimeric evolution in Fig. \ref{snapp}; Absolute values $L_i$ of the local curvature obtained through discrete Laplacian $l_i$ on the data of Fig. \ref{snapp}(a) and (b) are respectively shown in (b) and (c) here. }
	\label{af}
\end{figure}

 For a better perception on the dynamics of the system while exhibiting the chimera state, we plot the spatiotemporal evolution ($t \in [2000,~3000]$) in Fig. \ref{af}(a) with all the parameters kept fixed as in Fig. \ref{snapp}. From the figure, one is able to see the periodic (regular) repetition of the chimera profiles after certain time that points to the manifestation of a periodic alternating chimera pattern. Next we apply the discrete Laplacian (cf. Eq. (\ref{eq2})) on the snapshot of the membrane potentials $x_i$ (at time $t=1600$) as in Fig.\ref{snapp}(a) and obtain the profile of absolute value $L_i$ of the local curvature, shown in Fig. \ref{af}(b). The profile illustrates the spatial concurrence of coherence and incoherence from another aspect here. Similarly, $L_i$ applied on the snapshot at $t=1700$ (cf. Fig. \ref{snapp}(b)) is figured out in Fig. \ref{af}(c) the profile of which is having a contrasting nature to the previous one.

\begin{figure}
	\centerline{
		\includegraphics[scale=0.60]{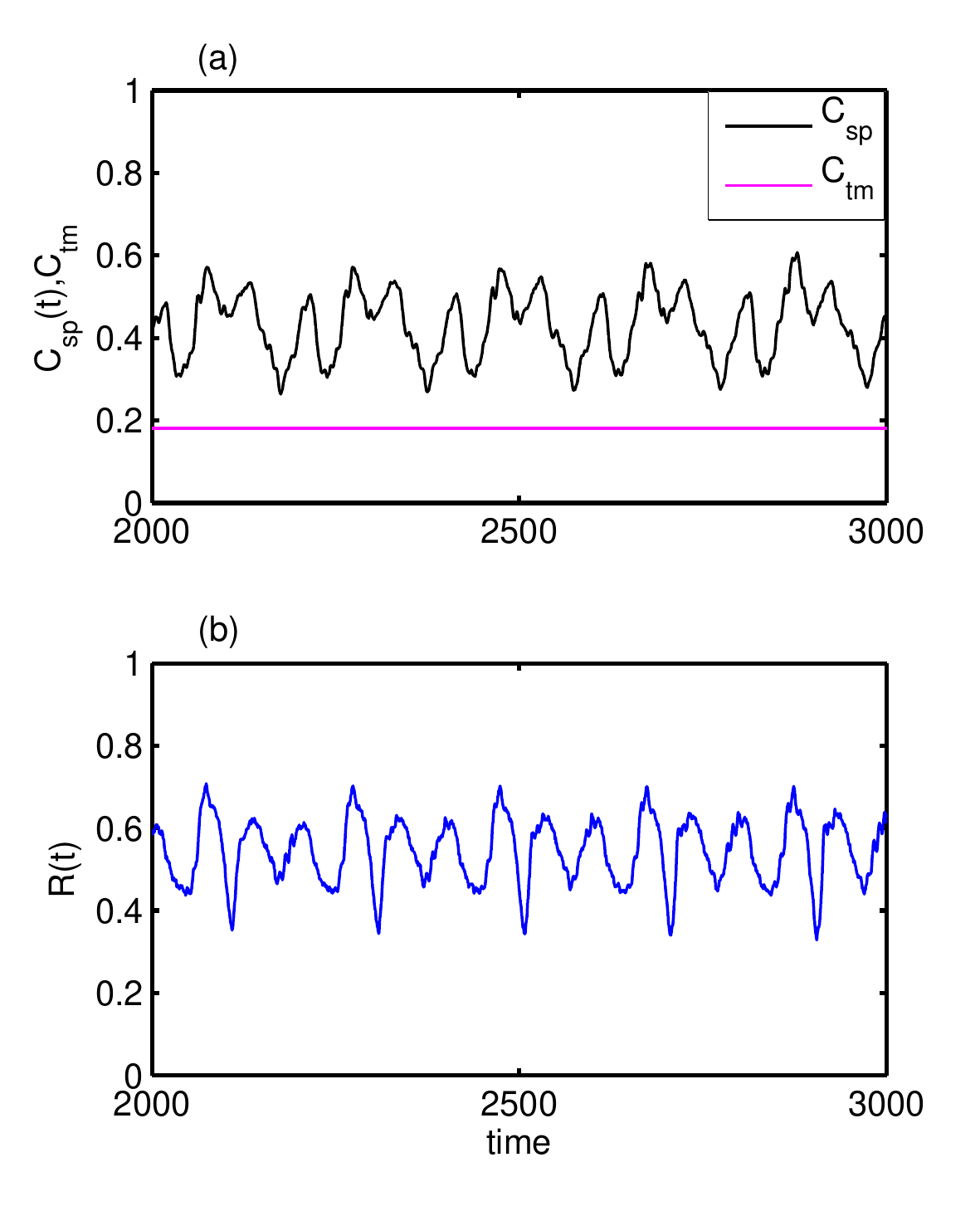}}
	\caption{(a) Spatial correlation measure $C_{sp}(t)$ as a function of time and the time-correlation measure $C_{tm}$ calculated over time interval $t \in [2000,~3000]$. (b) The order parameter $R(t)$  characterizing alternating nature of the chimera state. }
	\label{gh}
\end{figure}

\par Figure \ref{gh}(a) shows the spatial correlation measure $C_{sp}(t)$ as a function of time ($t \in [2000,~3000]$) where all the parameters are fixed same as Fig. \ref{af}(a). Periodic change in $C_{sp}(t)$ over time can be easily observed and this variation elaborates the emergence of the ``alternating chimera" pattern in the ephaptically coupled neuronal network. The time correlation measure $C_{tm}$ calculated over the entire time interval $t \in [2000,~3000]$ is found to be $C_{tm}\simeq 0.18$ which is also shown in Fig. \ref{gh}(a) implying that the alternating chimera is of stationary character. For further validation of the observed result on the alternating chimera, we compute the order parameter $R(t)$ as
\begin{equation}
\begin{array}{lcl}\label{eq7}
R(t)=|\frac{1}{N}\sum\limits_{j=1}^N e^{i \psi_j}|,
\end{array}
\end{equation}
where $i=\sqrt{-1}$ and the phase $\psi_j(t)$ is calculated in $(x_j, y_j)$ plane as  $\psi_j(t)=\mbox{tan}^{-1}[\frac{y_j(t)}{x_j(t)}]$ for $j=1,2,...,N$. This order parameter $R(t)$ basically quantifies the level of synchrony present in the system at the time instant $t$. Looking at the oscillating $C_{sp}(t)$ characterizing the alternating chimera, one expects a similar kind of behavior in the order parameter $R(t)$ as well. Figure \ref{gh}(b) depicts the variation in $R(t)$ with respect to time, that oscillates and consequently the chimera breathes. Regarding this nature of $R(t)$, the oscillating behavior in the order parameter is also observed previously in networks of phase oscillators with two small populations \cite{oscorder1}. Formation and destruction of traveling fronts \cite{oscorder2} in periodically modulated neural field model could also lead to oscillations in the order parameter. But, here, the regular alternating coexistence of coherence and incoherence is quite conspicuous from Fig. 2 and Fig. 3 (b, c) that demonstrate the origination of alternating chimera patterns which is valid even for large network sizes (cf. Appendix section).
\par
\begin{figure}
	\centerline{
		\includegraphics[scale=0.53]{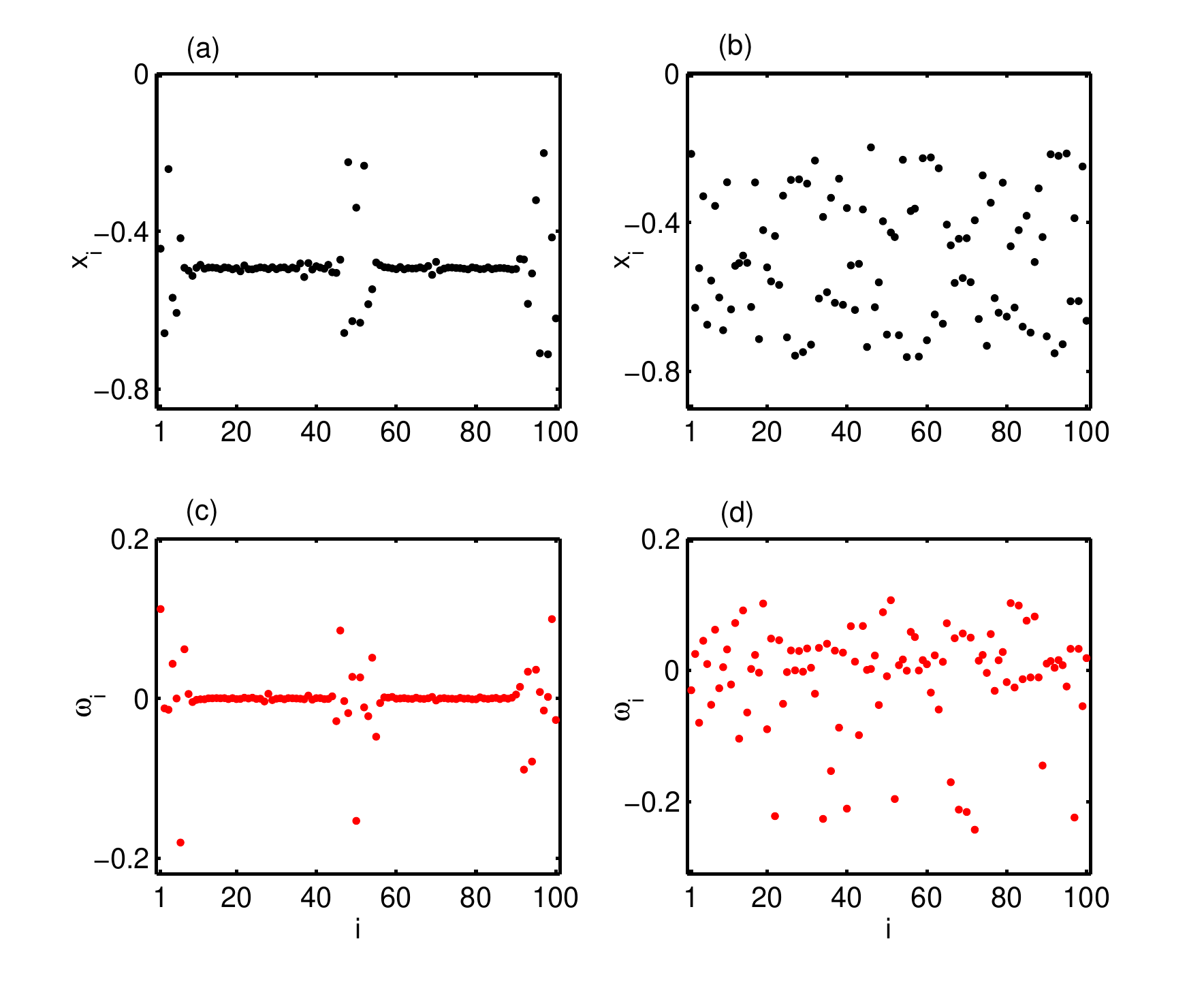}}
	\caption{ Snapshot of the membrane potentials $x_i$ reflecting (a) chimera and (b) incoherent states at $t=1600$ and $t=2500$ respectively. The corresponding instantaneous angular frequencies $\omega_i$ are respectively plotted in (c) and (d). The coupling strength $\epsilon=2.45$ and $r=0.01$ are chosen here. }
	\label{snp}
\end{figure}

 So far, we have concentrated on the dynamical behavior of the non-locally coupled neuronal network for a single value of $P=30$ (i.e., $r=0.3$) and we have found persistent (stationary) alternating chimera for a fixed interaction strength $\epsilon=0.5$. Next we focus on the local (nearest neighbor) configuration of the network in which $P=1$ ($r=0.01$) for which, of course, lower values of the coupling strength $\epsilon$ restrain the neuronal ensemble in the incoherent (disordered) state. This is true only up to $\epsilon=2.45$, beyond which the network starts experiencing chimera-like state. For instance, an exemplary snapshot of the membrane potentials for $\epsilon=2.45$ at the time $t=1600$ is shown in Fig. \ref{snp}(a) from which coexistence of coherence and incoherence can be realized. But, as one looks into a snapshot at higher time $t=2500$, this chimera profile completely dies out and rather the network behaves in a disordered fashion, as depicted in Fig. \ref{snp}(b). This implies the emergence of a typical transient chimera in the neuronal network. This circumstance is then identified by calculating instantaneous angular frequencies $\omega_i$ particularly at these times in Figs. \ref{snp}(c) and (d) respectively. The profiles of $\omega_i$ are quite self-explanatory as in the chimera state, the coherent clusters have similar frequencies while possessing no correlations in the incoherent domains.

\begin{figure}
	\centerline{
		\includegraphics[scale=0.55]{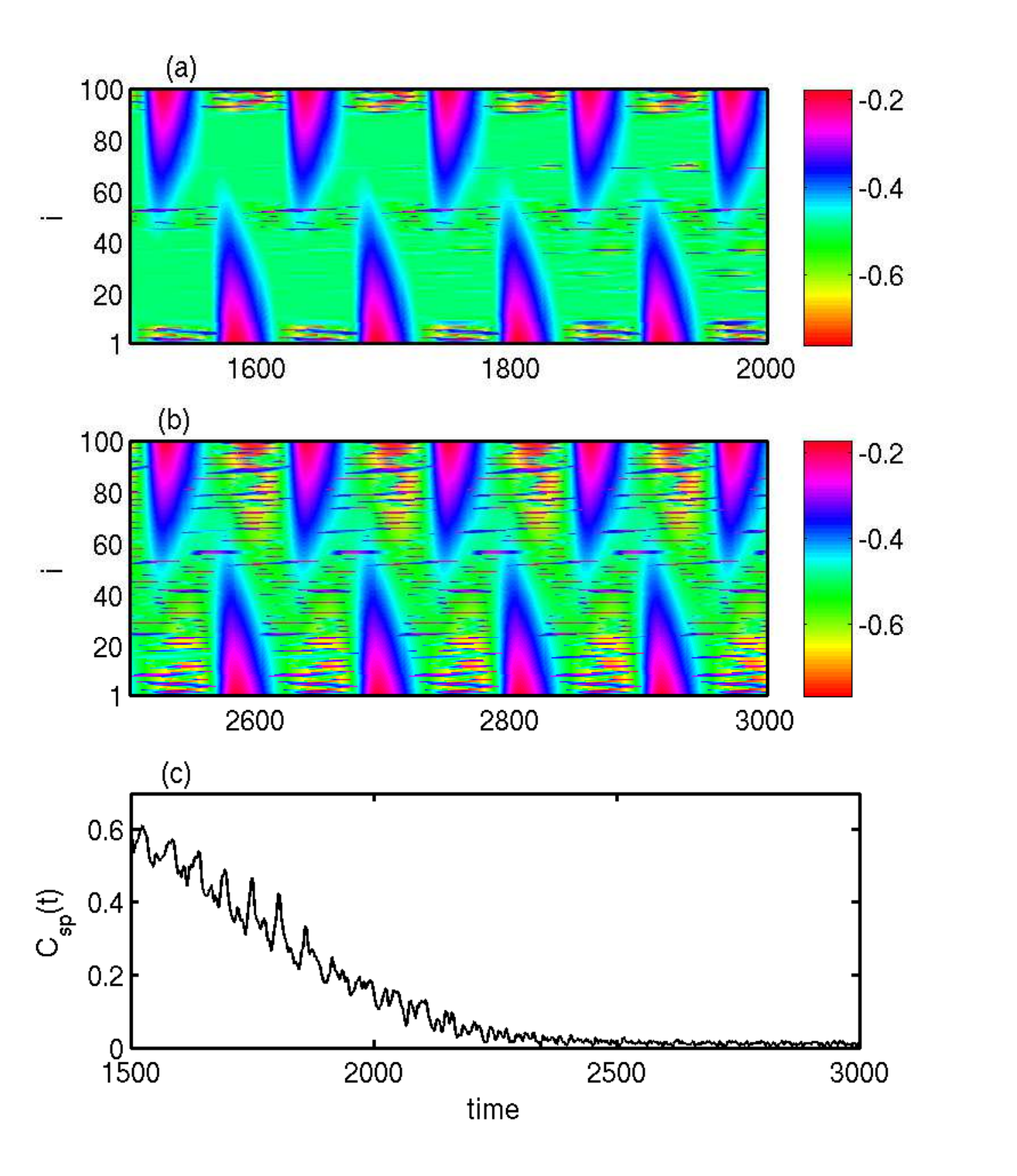}}
	\caption{ Spatio-temporal evolution over the time range (a) $t \in [1500,~2000]$ and (b) $t \in [2500,~3000]$ for $\epsilon=2.45$ with $r=0.01$; (c) Spatial correlation measure $C_{sp}(t)$ as a function of time over the interval $t \in [1500,~3000]$.}
	\label{pkch}
\end{figure}

\par This impermanence in the chimera state is thereafter justified by plotting the spatiotemporal evolutions of the network for two different time ranges $t \in [1500,~2000]$ and $t \in [2500,~3000]$ in Figs. \ref{pkch}(a) and (b) respectively for the same $\epsilon=2.45$. From Fig. \ref{pkch}(a), one can relate the scenario of chimera to the snapshot obtained earlier (cf. Figs. \ref{snp}(a) and (c)). However, there is no continuity of this phenomenon (of chimera) in the spatio-temporal plot of Fig.\ref{pkch}(b) plotted in higher time limit. Moreover, Fig. \ref{pkch}(c) shows variation in the spatial correlation measure $C_{sp}(t)$ that decreases over time and eventually happens to be zero. This feature of the transient chimera temporally leading to incoherence is in contrast to the observation reported in \cite{trnst,chhn1} in which a transition to full coherence was identified. These observations are further confirmed for larger size of the network, namely $N=500$ neurons by taking nonlocal ($P=150$) and local $(P=1)$ interactions [results are illustrated in Appendix].
\par
\begin{figure}
	\centerline{
		\includegraphics[scale=0.450]{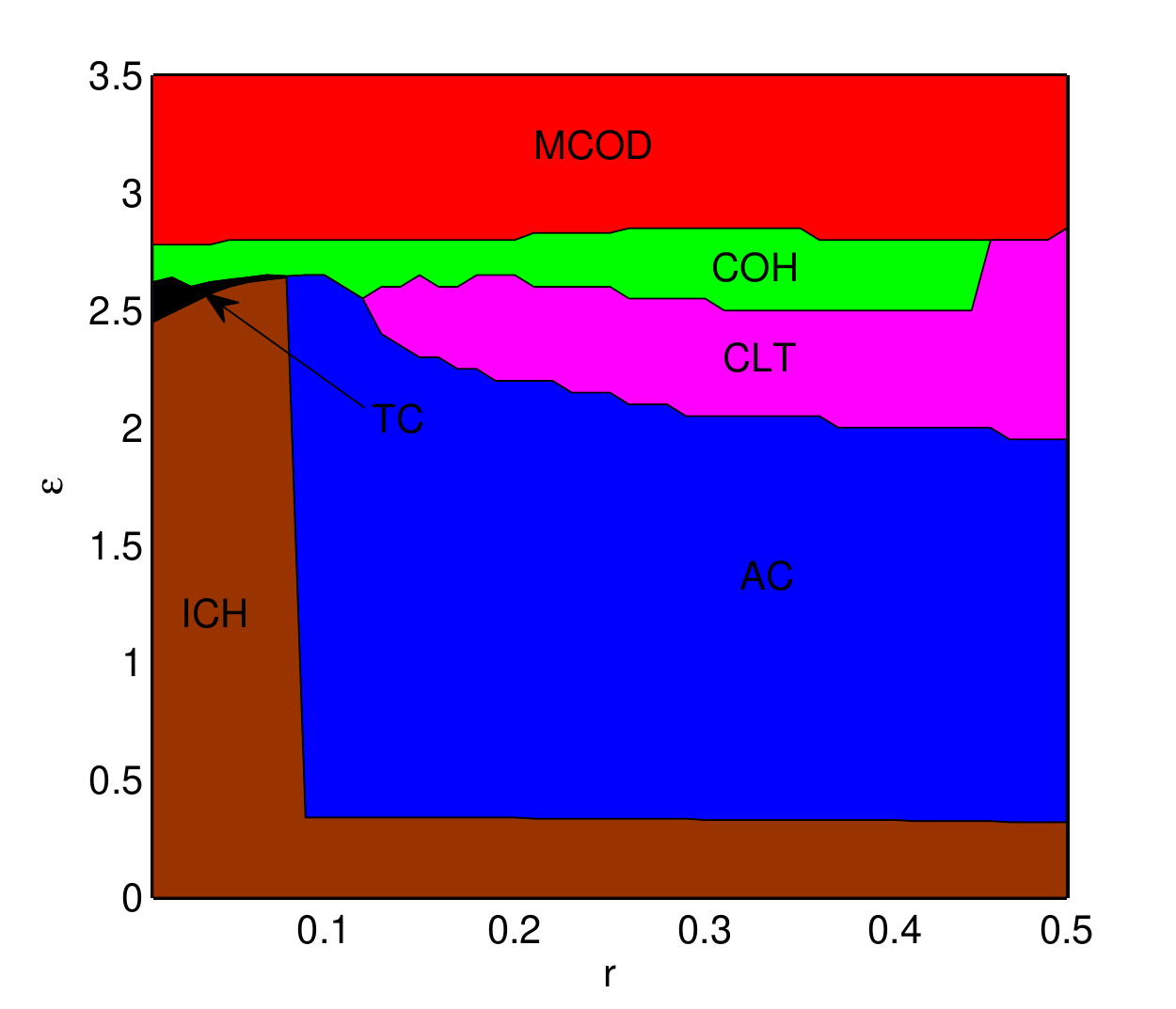}}
	\caption{ Two parameter phase diagram in $\epsilon-r$ plane where brown, blue, magenta, black, green and red colors respectively correspond to the incoherent (ICH), alternating chimera (AC), cluster (CLT), transient chimera (TC), coherent (COH) and multicluster oscillation death (MCOD) states.}
	\label{er}
\end{figure}

 This way we have realized stationary alternating chimera and transient chimera patterns in the network for $r=0.3$ and $r=0.01$ respectively, i.e. with non-local and local configurations. Now the time is for scrutinizing what is actually happening for other values of $r$ and what role is $\epsilon$ playing as far as the possible emanation of other dynamical states is concerned. In order to reveal this, we rigorously plot the phase diagram in the $\epsilon-r$ parameter plane in Fig. \ref{er} while keeping the other parameters fixed as before. 
In order to produce this phase diagram, we discriminate different dynamical states in the following way: for each and every value of $\epsilon$ and $r$, first we calculate the `$D$ factor', from which we can say whether the system exhibits oscillatory behavior or a steady state. We define the factor $D$ as
    \begin{equation}
	\label{eq:dfactor}
	D=\frac{1}{N}\sum_{i=1}^{N}\Theta\bigg(\sum_{l=1}^{T}|x_{i,t_l}-x_{i,t_{l-1}}|-\delta_3\bigg),
	\end{equation}
	where ${x_{i,t_l}}$, ${x_{i,t_{l-1}}}$ are the states of the ${i^{th}}$ oscillator at time iteration $t_l$ and its previous iteration $t_{l-1}$ with $T$ sufficiently large time and $\Theta(x)$ being the Heaviside step function. If the system goes to a steady state, value of $D$ will be `$0$'and if there is any sort of oscillation, then $D$ assumes value unity (i.e., `$1$') for proper choice of $\delta_3$. Whenever $D=1$, then we go for computing the values of $C_{sp}(t)$ and $C_{tm}$ over a sufficiently high time range. In fact, $C_{sp}(t)\simeq 0 ~\big(C_{sp}(t)\simeq 1\big)$ represents incoherence \big(coherence\big), the periodically repeating values with $0<C_{sp}(t)<1$ signifies the regular alternating chimera. Also, the scenario of $0<C_{sp}(t)<1$ decaying to zero implies the transient chimera and in our work, any constant $C_{sp}(t)$ such that $0<C_{sp}(t)<1$ stands for stable cluster states here. Finally, the regions in the phase diagram with $D=0$ represents oscillation suppression state that happens to be the multicluster oscillation death state here. We have fixed the value $\delta_3=0.005$. 
\par In the limit of local coupling, particularly for $0.01 \le r \le 0.08$ the network retains its disordered state even upto $\epsilon=2.45$, above which transient chimera may appear (depending on the value of $r$) as a link between incoherence and coherence followed by multicluster oscillation death (MCOD) state \cite{mcod}. Although a different picture of transition is observed with $0.09 \le r \le 0.49$ for which the alternating chimera states appear quite early at $\epsilon \ge 0.33$ and may persist up to maximum $\epsilon=2.60$ (depending on $r$). Beyond this value, the network experiences coherent and MCOD states respectively for increasing $\epsilon$. Interestingly, for the coupling range $0.12< r \le 0.49$, in addition to the earlier noted states, synchronized cluster (CLT) states \cite{clt} appear in between the alternating chimera and coherent patterns. But for the rest of $r \in [0.45,~0.49]$ (even for global interaction), MCOD states emerge right after the appearance of cluster states for increasing $\epsilon$.

\subsection{Neurons exhibiting periodic square-wave bursting}

\begin{figure}
	\centerline{
		\includegraphics[scale=0.50]{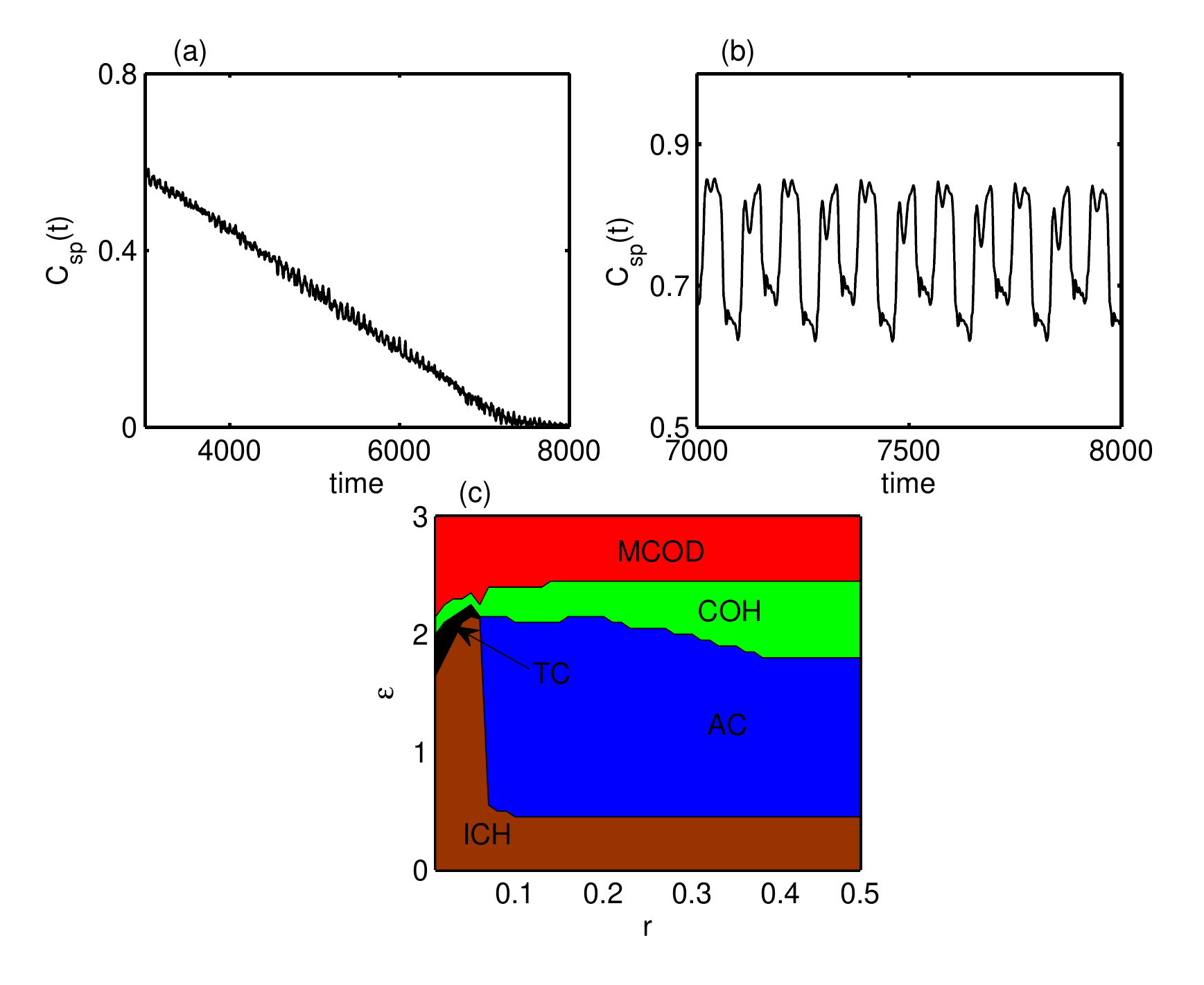}}
	\caption{  Spatial correlation measure $C_{sp}(t)$ characterizing (a) transient chimera (for $r=0.01$ and $\epsilon=1.90$) and (b) alternating chimera (for $r=0.3$ and $\epsilon=0.45$) states; (c) Phase diagram in $\epsilon-r$ coupling parameter plane where brown, blue, black, green and red colors respectively correspond to the incoherent (ICH), alternating chimera (AC), transient chimera (TC), coherent (COH) and multicluster oscillation death (MCOD) states.}
	\label{er22}
\end{figure}

Whenever neurons in the network are inferred to exhibit regular (periodic) square-wave bursting dynamics (cf. Fig. \ref{tsr}(b)) with the external stimulus $I=1.90$, a more or less similar qualitative phenomenon has been witnessed. As in the earlier observation for chaotic square-wave bursting of the neurons with $N=100$, here also for lower coupling radius ($r \in [0.01,~0.06]$) the network goes through transient chimera patterns as a link between incoherence and coherence followed by MCOD states. As an exception, now the transient chimera states may have higher lifetime, as depicted by the spatial correlation measure $C_{sp}(t)$ over time $t \in [3000,~8000]$ in Fig. \ref{er22}(a) with $r=0.01$ and $\epsilon=1.90$. The chimera states obtained for non-local coupling ($r \in [0.07,~0.49]$) ensue a stationary alternating characteristic, an exemplary oscillating $C_{sp}(t)$ profile is shown in Fig. \ref{er22}(b) for $r=0.3$ and $\epsilon=0.45$. In Fig. \ref{er22}(c), we figure out the transition scenario of the dynamical network for simultaneous variation in the coupling parameters $\epsilon \in [0,~3]$ and $r$. As pointed out, with the coupling radius $r \in [0.01,~0.06]$, transient chimera arises whereas for all the other possible values of $r$ alternating chimera appears. But in contrast to the chaotic bursting case, here the neurons become directly coherent from the state of alternating chimera as a result of increasing coupling strength without experiencing the cluster states and finally reaches the oscillation quenching state (MCOD).

\subsection{Plateau bursting of the neurons}

\begin{figure}
	\centerline{
		\includegraphics[scale=0.52]{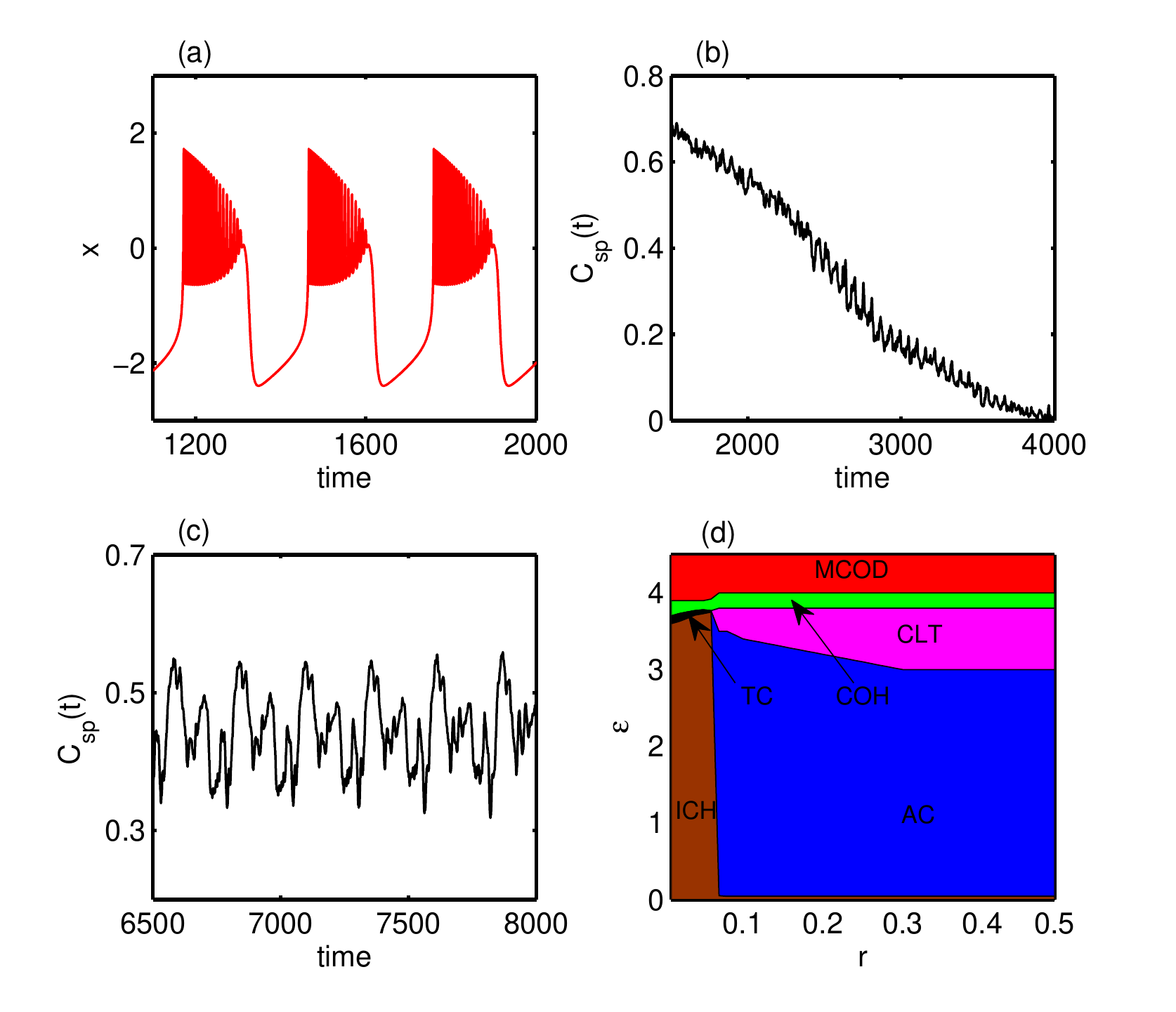}}
	\caption{(a) Plateau bursting dynamics of the neurons with $I=3.0$ and $b=2.50$; Spatial correlation measure $C_{sp}(t)$ characterizing (b) transient chimera (for $r=0.01$ and $\epsilon=3.60$) and (c) alternating chimera (for $r=0.3$ and $\epsilon=0.20$) states; (d) Phase diagram in the $\epsilon-r$ parameter plane with brown, blue, black, magenta, green and red colors respectively depicting the incoherent (ICH), alternating chimera (AC), transient chimera (TC), cluster (CLT), coherent (COH) and multicluster oscillation death (MCOD) states.}
	\label{er32}
\end{figure}

In this subsection, we evidence the emergence of alternating chimera states even when the neurons in the ensemble ensue a different sort of bursting dynamics, namely plateau bursting.
Exemplary time evolution of the membrane potentials corresponding to plateau bursting is shown in Fig. \ref{er32}(a). In this context, for the local coupling limit, transient chimera pattern arises within a very narrow range of the interaction strength $\epsilon$. For instance, with $r=0.01$ and $\epsilon=3.60$, the chimera states observed are essentially short-lived (in time). This dynamical feature of the network is explained in terms of decaying profile of the spatial correlation measure $C_{sp}(t)$ within the time interval $[1500,~4000]$ in Fig. \ref{er32}(b). Besides, non-local coupling induces periodically alternating nature in the chimera patterns even if the interaction strength is very small. Whenever the coupling radius $r$ lies in the range $[0.07,~0.49]$, alternating chimera starts appearing for $\epsilon \ge 0.05$. Periodically pulsating $C_{sp}(t) \in (0,~1)$ profile identifying alternating chimera has been presented in the Fig. \ref{er32}(c) for $t \in [6500,~8000]$ where $r=0.3$ and $\epsilon=0.2$. As the next step, phase diagram in the $\epsilon-r$ coupling parameter plane is plotted in Fig. \ref{er32}(d) for a complete understanding of the impacts of these parameters on the network. As stated earlier, within the coupling radius $0.01 \le r \le 0.06$, the network realizes chimera pattern which is impermanent and leads to incoherence over time (cf. black region in the figure). For these coupling radii, with increasing $\epsilon$, the network experiences coherent state followed by MCOD state. For higher coupling radius $r \in [0.07,~0.49]$, the alternating chimera patterns occur in a much larger portion of the parameter plane than the previous two cases (as shown by blue color in the figure), appearing for the maximum range $0.05 \le \epsilon \le 3.50$ depending on $r$. For $\epsilon$ beyond this range, cluster synchronization arises in the ensemble similarly as in case of chaotic square-wave bursting neurons. Coherent states appear with higher interaction strength $\epsilon$ followed by the MCOD state. One more important thing this figure explicitly shows is the prolongation of the alternating chimera region in the $\epsilon-r$ parameter plane. In this sense, the plateau bursts promote alternating chimera pattern more than the other two bursting dynamics considered here.

\section{Conclusions}
The concurrence of dynamical coherence and incoherence in a symmetrically configured network has strong resemblance to several neuronal developments
and so far networks of neurons based upon synaptic communication have been treated to demonstrate possible appearance of such chimera-like states. But synaptic coupling is not the only interacting medium among the neurons, rather external fields in the form of magnetic flux across the membrane can act as a supplementary mode of information exchange. So in contrast, in this work we have presented the emergence of chimera patterns in a network of neurons in absence of any sort of synaptic connectivity among them. Specifically, neurons are assumed to be connected through ephaptic coupling of nerve fibers by virtue of local electric fields. Whenever chaotic bursting dynamics arising from Hindmarsh-Rose neuron model has been used to cast the local dynamics of the nodes, stationary alternating chimera is realized for both non-local and global interactions. However, transient chimeras are identified for local coupling. Besides these dynamical states, coherent patterns along with cluster and multicluster oscillation death states are also observed in the network depending on the values of coupling radius and interaction strength. Nevertheless, strong coupling induces multicluster oscillation death state irrespective of the value of coupling radius. For periodic bursting of the neurons, the lifetime of the transient chimera states may increase. Moreover increasing interaction strength guides the network towards the coherent state directly from  state of alternating chimera without going through cluster states, if the neurons interact non-locally or globally. We have validated our results on obtained dynamical phenomena for neurons exhibiting plateau bursting dynamics and also observed that this plateau bursting actually expands the alternating chimera region in the coupling parameter plane quite comprehensively.
\par  We would like to further mention that the alternating chimera patterns in our work are realized in a network of indirectly coupled neuronal systems through electromagnetic field whereas the previous works \cite{chp3,chtv,chyao}, on the same phenomena were mainly investigated in directly connected networks of phase oscillators. Besides, the transient chimera persist irrespective of the neuronal network size and remarkably this transitory feature appears even in purely locally interacting ensemble. The acquired outcomes thus raise the significance of ephaptic communication in neuronal ensembles from the perspective of several dynamical consequences and further relaxes the requirement of synaptic communication in order to experience chimera patterns in the network.

\section{Appendix}

\begin{figure}
	\centerline{
		\includegraphics[scale=0.55]{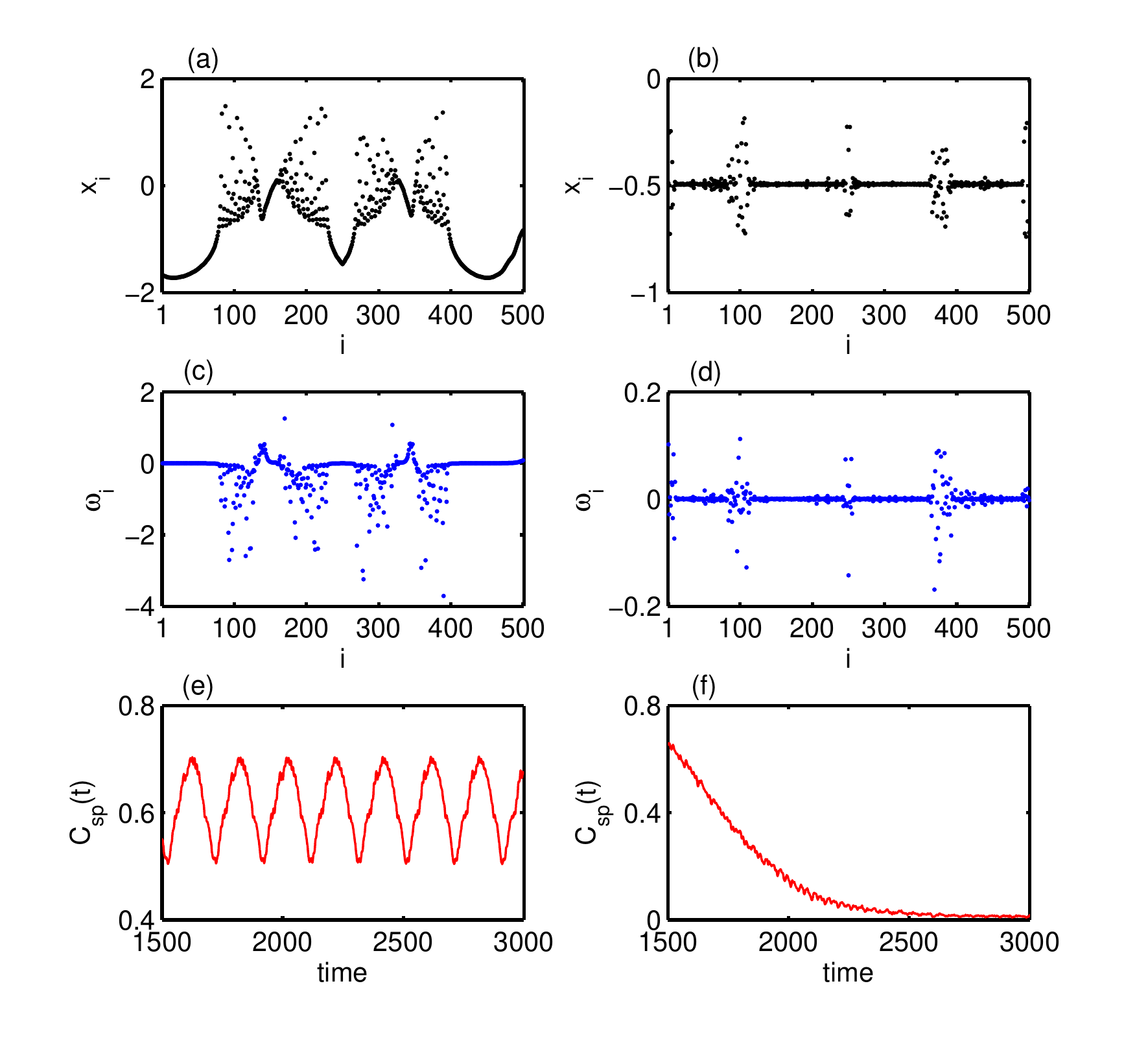}}
	\caption{ (a,b) Snapshot of the membrane potentials at $t=1500$ showing chimera patterns for $P=150$, $\epsilon=0.5$ and $P=1$, $\epsilon=2.45$ respectively. Corresponding instantaneous angular frequencies are shown in (c,d). Spatial correlation measure $C_{sp}(t)$ reflecting alternating and transient natures of chimera states are respectively plotted in (e) and (f). }
	\label{gl}
\end{figure}

In order to demonstrate that the observed qualitative results on the alternating (non-local interaction) and transient (local coupling limit) chimeras do not depend on the network size and persist even for larger networks, we went for analyzing the possible network behaviors for non-local and local interactions in which the number of neurons in the network is $N=500$. For the sake of simplicity, we kept all other parameters fixed as above (in case of $N=100$). Figure \ref{gl}(a) shows snapshot of the membrane potentials $x_i$ at $t=1500$ that depicts coexistence of coherence and incoherence for $r=0.3$ with $\epsilon=0.5$. On the other hand, with $r=0.002$ (local coupling) and $\epsilon=2.45$, snapshot at the same time-instant is plotted in Fig. \ref{gl}(b) that also depicts chimera state. In addition, as a confirmation of these states, we have shown the angular frequencies $\omega_i$ corresponding to these snapshots in Figs. \ref{gl}(c) and (d) respectively. To have a perception about the lifetime of these chimera patterns, we further plot the spatial correlation measure $C_{sp}(t)$ in Figs. \ref{gl}(e) and (f). Figure \ref{gl}(e) shows (periodically) pulsating $C_{sp}(t)$ throughout $t \in [1500,~3000]$ indicating an alternating nature of the chimera state whereas it eventually drops down to zero signifying a transient behavior, as in Fig. \ref{gl}(f).\\
\\
\begin{acknowledgments}
    D.G. was supported by SERB-DST (Department of Science and Technology), Government of India (Project no. EMR/2016/001039).
\end{acknowledgments}


\begin{thebibliography}{25}
\bibitem{gene} We opted these particular values, however, small parameter variations do not affect the observed qualitative results.
\bibitem{mthdinit} The fifth-order Runge-Kutta-Fehlberg integration scheme is used to integrate the system (1) with integration time step 0.01. The initial conditions (general V-shaped) are chosen as: $x_i(0)=a_1(N/2-i)$, $y_i(0)=a_2(N/2-i)$, $z_i(0)=a_3(N/2-i)$ for $i=1,2,...,N/2$ along with $x_i(0)=b_1(i-N/2)$, $y_i(0)=b_2(i-N/2)$, $z_i(0)=b_3(i-N/2)$ for $i=N/2+1,...,N$ where $a_1=0.01,~a_2=0.02,~a_3=0.03$ and $b_1=0.012,~b_2=0.024,~b_3=0.035$. However, there are no significant changes in the occurrence of the alternating chimera for different choices of these coefficients that satisfies $x_i(0) \in [-5,~5],~y_i(0) \in [-10,~2]$ and $z_i(0) \in [0,~3]$.
\end{thebibliography}
\end{document}